\documentclass[11pt]{article}
\usepackage{theorem}
\usepackage{latexsym}
\usepackage{amsmath}

\usepackage[dvips]{graphicx}
\usepackage{graphicx}

\theoremheaderfont{\scshape}
\newtheorem{theorem}{Theorem}[section]
\newtheorem{corollary}{Corollary}[section]
\newtheorem{lemma}{Lemma}[section]
\newtheorem{proposition}{Proposition}[section]
\newtheorem{postulate}{Postulate}[section]
\newtheorem{statement}{Statement}[section]
{\theorembodyfont{\rmfamily}
\newtheorem{example}{Example}[section]}
{\theorembodyfont{\rmfamily}
\newtheorem{remark}{Remark}[section]}
{\theorembodyfont{\rmfamily}
\newtheorem{definition}{Definition}[section]}

\newcommand{\newc}{\newcommand}

\newc{\be}{\begin{equation}}
\newc{\ee}{\end{equation}}
\newc{\bea}{\begin{eqnarray}}
\newc{\eea}{\end{eqnarray}}
\newc{\beas}{\begin{eqnarray*}}
\newc{\eeas}{\end{eqnarray*}}

\newc{\pardt}{\partial_{t}}
\newc{\pardxi}{\partial_{i}}
\newc{\pardts}{\partial_{t^{*}}}
\newc{\pardxis}{\partial_{i^{*}}}
\newc{\pardxj}{\partial_{j}}
\newc{\pardxk}{\partial_{k}}
\newc{\pard}{\partial}

\newc{\s }{\overline}

\newc{\sect}{\section}
\newc{\subs}{\subsection}

\newc{\defi}{\definition}
\newc{\prop}{\proposition}
\newc{\rem}{\remark}
\newc{\lem}{\lemma}
\newc{\exa}{\example}
\newc{\theo}{\theorem}
\newc{\coro}{\corollary}
\newc{\post}{\postulate}
\newc{\state}{\statement}

\begin{document}
\baselineskip0.5cm
\renewcommand {\theequation}{\thesection.\arabic{equation}}
\title{Waves propagation in turbulent superfluid
helium\\ in presence of combined rotation and counterflow}
\author{ R. A. Peruzza \and M.
Sciacca\thanks{Corresponding author. } }

\date{ }
\maketitle
\begin{center}{\it Dipartimento di Metodi e Modelli Matematici
Universit\`a di Palermo, \\ Facolt\`a di Ingegneria, Viale delle
Scienze, 90128 Palermo, Italy} \vskip.5cm
\end{center}
\footnotetext{E-mail addresses: peruzza@unipa.it (R.A. Peruzza),
msciacca@unipa.it (M. Sciacca) }

\begin{abstract}
A complete study of the propagation of waves (namely longitudinal
density and temperature waves, longitudinal and transversal
velocity waves and heat waves) in turbulent superfluid helium is
made in three situations: a rotating frame, a thermal counterflow,
and the simultaneous combination of thermal counterflow and
rotation. Our analysis aims to obtain as much as possible
information on the tangle of quantized vortices from the wave
speed and attenuation factor of these different waves, depending
on their relative direction of propagation with respect to the
rotation vector.
\\
\\
\small {{\it PACS number(s)}: 05.70.Ln, 67.40.Vs, 67.40.Bz,
47.27.2i\\ {\it A.M.S. Classification}: 82D50\\
{\it Key words}: non equilibrium thermodynamics; superfluid
turbulence; second sound; vortex tangle}

\end{abstract}

\section{Introduction}
The most known phenomenological model, accounting for many of the
properties of He II, given by {\it Tisza} \cite{Tisza} and {\it
Landau} \cite{Landau} is called the {two-fluid model}. The basic
assumption is that the liquid behaves as a mixture of two fluids:
the normal component with density $\rho_n$ and velocity ${\bf
v}_n$, and the superfluid component with density $\rho_s$ and
velocity ${\bf v}_s$. When the difference ${\bf V}:={\bf v}_n-{\bf
v}_s$ between the normal and superfluid velocities, known also as
counterflow velocity, exceed a certain critical velocity, a {\it
mutual friction} ${\bf F}_{sn}$ has to be included. This friction
force is attributed to an interaction of the normal component with
the vortices in the superfluid.
\par Quantized superfluid vortices play an important role in the
hydrodynamics of the fluid and they have been the object of many
studies. The state of the fluid in which vortices are present, is
referred to as the {\it superfluid turbulent state}. A review on
superfluid turbulence can be found in Tough's paper \cite{Tough}
and in chapter 7 of Donnelly's book \cite{Donnelly}. The quantized
vortices created by applying a thermal counterflow form an
irregular, spatially disordered tangle of lines. In this case, the
vortex line density $L$ (length of vortex line per unit volume) is
$L_H\approx\gamma^2V^2$, where ${\bf V}$ is the modulus of the
relative velocity between the two components of the mixture and
$\gamma$ a temperature-dependent coefficient \cite{Tough}. The
vortex system is almost isotropic, provided that one neglects a
small anisotropy induced by the imposed counterflow \cite{WSD}.
\par The creation of the vortices cannot be made only in this
way; in fact, the first studies of quantized vorticity involved a
sample of He II rotating at constant angular velocity $\Omega$
exceeding a certain small critical value. The results brought to
an ordered array of vortices aligned along the rotation axis,
whose number density per unit area is given by Feynman's rule
$L_R=2\Omega /\kappa$, where $\kappa=h/m=9.97\quad\!\!\! 10^{-4}$
cm$^2/$sec is the quantum of circulation, with $h$ Planck's
constant and $m$ mass of the helium atom.
\par Now, an important question naturally arises: what happens if
vortices are created by both rotation and counterflow? There has
been only one experiment of which we are aware \cite{SBD}, on the
formation of vortices in combined rotation and counterflow along
the rotational axis. This experiment suggests that there exists a
form of steady rotating turbulence, characterized by a vortex line
density at given counterflow velocity ${\bf V}$ and angular
velocity ${\bf \Omega}$. Swanson et al. \cite{SBD} found that at
slow rotation the critical counterflow velocity above which the
flow became turbulent was greatly reduced. The experimental
observations showed that the two effects (thermal counterflow and
rotation) are not merely additive, in fact for $V $ high the
measured values of $L$ are always less than $L_H+L_R $. However,
from our point of view the results of these experiments are purely
qualitative because the authors didn't take the anisotropy of the
vortex tangle in consideration which, as we will see in the last
section of this paper, is essential to know the spatial
distribution of the vortex tangle in liquid Helium II through
measurements of second sound attenuation.

The aim of this work is to study the propagation of longitudinal
density and temperature waves,  and longitudinal and transversal
velocity waves and heat waves in the combined situation of a
rotating frame and of a cylindrical container in presence of
thermal counterflow. The studies of the two separate cases of pure
rotation and pure thermal counterflow are also considered in order
to give a more complete view of the wave propagation in these
three different situations. The influence of the parameters
characterizing the vorticity on the propagation of the waves is
shown explicitly. The practical interest of this research is  to
obtain information on the vortex tangle from measurements on wave
propagation. This is an important issue, because under the
combined influence of rotation and counterflow the vortex tangle
cannot be assumed isotropic. Then, we must find not only the
vortex line density $L$ but also the geometrical characterization,
which requires, in principle, to consider wave propagation in
different directions, as well as a deeper full analysis of waves.
Note that here the tangle itself is not considered as a dynamical
quantity, because it is not modified by the second sound. For this
reason, evolution equations for the tangle are not needed here.
\par\noindent The plan of this paper is the following: Section 2
is concerned with the model for helium II, in which the use of a
pressure tensor associated to the vorticity has been considered;
in Section 3 and in Section 4 we study wave propagation in
rotating frame and in presence of thermal counterflow
respectively, pure rotation is analyzed in the general case in
which a component of the mutual friction force parallel to the
rotation axis is present; finally, in Section 5 we study wave
propagation in simultaneous rotation and counterflow, analyzing
two different situations about the relative direction of wave
propagation with respect to the rotation vector.

\section{Evolution equations}
Many observations have shown  that both thermal conductivity
$\lambda_1$ and the relaxation time of the heat flux $\tau_1$ in
helium II are very high. As observed in \cite{M93} their ratio
$\frac{\lambda_1}{\tau_1}:=\zeta<\infty$ determines the velocity
of the second sound, which is a heat wave propagating in the
superfluid. As a consequence, it is natural to use a
thermodynamical theory where the heat flux $\bf q$ appears as a
further fundamental field. In this way, a linear macroscopic
one-fluid model of liquid helium II, based on Extended
Thermodynamics \cite{JCL, MR}, has been formulated \cite{M93}.
This model is able to describe the laminar flow of the superfluid
both in the presence and in absence of dissipative phenomena and
to predict the propagation of the two sounds in bulk liquid helium
II and of the fourth sound in liquid helium flowing in a porous
medium \cite{M93}, \cite{MPz}-\cite{MPJ}, in agreement with
microscopic and experimental data.
\par
In order to describe the presence of vortices in rotating helium
II, in superfluid turbulence or in combined rotation and thermal
counterflow, the use of a further additional pressure tensor ${\bf
P}_\omega$, associated to the vorticity, is necessary. The
simplified situations of a rotating frame and of pure thermal
counterflow have been considered in \cite{JLM}, where a
constitutive relation for ${\bf P}_\omega$ and its influence on
the dynamics of the heat flux has been studied.

In this work the more complex situation involving thermal
counterflow in a rotating cylinder, which is receiving much
attention recently \cite{BDV}-\cite{Tsubota2004}, is considered
too. We start from a linear macroscopic one-fluid model of liquid
helium II, whose fundamental fields are the density $\rho$, the
barycentric velocity ${\bf v}$, which is related to the two
velocities of the two-fluid model by the relation $\rho {\bf
v}=\rho_s {\bf v}_s+\rho_n {\bf v}_n$, the temperature $T$ and the
heat flux ${\bf q}$, related to the counterflow velocity ${\bf V}$
by the relation ${\bf q}=\rho_s T s {\bf V}$ (where $s$ is the
entropy of the Helium II). In the two-fluid model the natural
variables are ${\bf v}_s$ and ${\bf v}_n$, but in the experiments
it is ${\bf v}$ and ${\bf q}$ which are directly measured.
Therefore, the use of ${\bf v}$ and ${\bf q}$ appears suitable for
our analysis. Neglecting the bulk and shear viscosity and under
the hypothesis of small thermal dilatation (which in helium II are
indeed very small), the linearized system of field equations for
liquid helium II, in a non inertial frame, in absence of external
force, is \cite{JLM}:
\begin{equation}
\begin{cases}
\frac{\partial \rho}{\partial t}+\rho\frac{\partial v_j}{\partial
x_j }=0\cr \rho\frac{\partial v_i}{\partial t}+\frac{\partial
p}{\partial x_i}+{\bf i}^0_i+2\rho\left({\bf \Omega}\wedge {\bf
v}\right)_i=0\cr \frac{\partial T}{\partial t}+\frac{1}{\rho
c_V}\frac{\partial q_j}{\partial x_j}=0\cr \frac{\partial
q_i}{\partial t}+\zeta\frac{\partial T}{\partial x_i}+2\left({\bf
\Omega}\wedge{\bf q}\right)_i=\left({\vec \sigma}_\omega\right)_i
=-\left({\bf P}_\omega\cdot{\bf q}\right)_i.
\end{cases}\label{1}
\end{equation}
 In this system, ${\bf i}^0+2\rho\left({\bf \Omega}\wedge {\bf v}\right)$ is the
inertial force, $\zeta$ is a positive coefficient linked to the
second sound velocity, and:
\begin{equation}
p=p_E(\rho,T)\quad \mbox{and}\quad c_V=\left(\frac{\partial \epsilon
(\rho ,T) }{\partial T}\right)_\rho
\end{equation}
are the thermostatic pressure and the specific heat respectively
($\epsilon$ is the specific internal energy). The effect of
vortices is described by  incorporating the source term ${\bf
P}_\omega\cdot{\bf q}$ to the evolution equation of the heat flux.
As we will see, the expression of  ${\bf P}_\omega$ will assume
different expressions in the different situations considered.

Now, a small comparison between the one-fluid model and the
two-fluid model could be useful. With the corresponding
transformations between the natural variables in the one fluid
model, ${\bf v}$ and ${\bf q}$, and those in the two-fluid model,
${\bf v}_s$ and ${\bf v}_n$, the evolution equations (\ref{1}) of
the one-fluid model are equivalent to those of the two-fluid model
in the linear approximation \cite{JLM}. A formal difference is
found in the form of the production term in the evolution equation
for the heat flux (\ref{1}d). When specified to pure rotation,
this production term, as given by (\ref{sigma2}), has the usual
Hall-Vinen form, whereas when specified to counterflow, the
production term, as given by (\ref{4.1}), yields the well-known
Gorter-Mellinck form. These two situations have been well explored
in the context of the two-fluid and one-fluid frameworks. In the
combined situation with simultaneous rotation and counterflow, the
general form of the production term in (\ref{1}d) is especially
useful, as expressed in (\ref{pesp2}) and (\ref{5.38}), because it
allows one to write in an explicit and appealing way the
anisotropy of the tangle, whose influence on the second sound is
one of our main concerns. Given the same geometrical conditions
for the tangle --- which here are given a priori, and whose form
is probed by means of second sound ---, the evolution equations of
the two-fluid model would coincide with those of the one-fluid
model. Thus, the dispersion relations obtained here should be
valid also in the context of the two-fluid model.

The one-fluid and the two-fluid models are not identical to each
other. However, their mutual differences arise in contexts which
are not relevant in the analysis presented here. For instance, one
difference arises in the fourth sound in helium through porous
media, in which some experimental results seem to support the
one-fluid model \cite{Mongiovi}. Anyway, the two-fluid model could
also cope with that situation provided the assumption that the
superfluid component carries no entropy is slightly relaxed by
assuming that it may carry a small but nonvanishing entropy. Other
differences arise concerning the interaction between second sound
and the vortex tangle. Here, we have assumed that second sound
does not modify the vortex line density nor the geometrical
structure of the tangle. If it is assumed that it may modify the
vortex tangle, more general evolution equations would be needed,
as for instance an evolution equation for the vortex line density
$L$ coupled with the rotation and counterflow, which have already
been explored in the literature \cite{MonJou07}.  For instance,
the evolution equation for $L$ could be different --- a
generalized form of Vinen's equation with the mentioned couplings
has been proposed and studied \cite{JM04} --- but this is not
relevant here because an equation for $L$ is not necessary in this
paper, as $L$ is taken as fixed, and its value must be found from
wave experiments. Some other differences may appear, concerning,
for instance, the possibility of vortex density waves at high
frequencies in the one-fluid model \cite{JMS} that do not arise in
the Hall-Vinen-Bekarevich-Khalatnikov model \cite{Henderson}.
Since in this work we are focusing our attention to a situation in
which the interaction between the second sound and the tangle does
not distort the vortex lines nor the vortex density, the
dispersion relations obtained in this paper by using the
production terms ${\bf P}_\omega \left({\bf q}, {\bf
\Omega}\right) \cdot{\bf q}$ would be also valid in the two-fluid
context by using a production term of the form ${\bf P}_\omega
\left({\bf v}_n-{\bf v}_s, {\bf \Omega}\right) \cdot ({\bf
v}_n-{\bf v}_s)$ in an evolution equation for the relative
velocity ${\bf V}={\bf v}_n-{\bf v}_s$.

\section{Wave propagation in rotating frame}
\setcounter{equation}{0}

We generalize here the results of \cite{JLM} to the case in which
a small interaction between second sound and vortex line parallel
to the rotation axis is present. In \cite{HV56}, Hall and Vinen
described experiments of liquid Helium II in a rotating frame,
showing the main effects on the propagation and attenuation of the
second sound as a consequence of the interaction between
quasi-particles and vortex lines: these interactions are mainly
present in the planes orthogonal to the rotation axis. As
consequence of these experiments, in \cite{JLM}, Jou, Lebon and
Mongiov\`{\i} proposed an expression for the production term
${\vec{\sigma} }_{\omega}$ in (\ref{1}d), which takes into account
dissipative and non dissipative contributions of the interaction
between quasi-particles and vortex lines, but they did not
consider interactions parallel to the rotation axis.

In another experiment \cite{Snyder}, Snyder studied the component
of mutual friction along the rotational axis, and his result, in
agreement with  \cite{BTMA2}, shows that this friction component
is very small compared with the orthogonal components but not
exactly zero. In this section, we consider the most general case
in which the axial component is included. In order to do that, the
following vorticity tensor ${\bf P_\omega}$ is used \cite{JM05}:
\bea \label{Pomega} {\bf P}_\omega^R=\frac{1}{2}\kappa L_R
\left[(B-B'')\left({\bf U}-{\bf \hat{\Omega} \hat{\Omega}}\right)+
B' {\bf W\cdot \bf \hat{\Omega}}+2B''\bf
\hat{\Omega}\hat{\Omega}\right],\eea where ${\bf U}$ is the unit
matrix, ${\bf W}$  the Ricci tensor, and $B$ and $B'$ are the
Hall-Vinen coefficients \cite{HV56} describing the orthogonal
dissipative and non dissipative contributions while $B''$ is the
friction coefficient along the rotational axis. Using the Eq.
(\ref{Pomega}), the production term in (\ref{1}d) can be expressed
as \cite{Donnelly,JM05}: \bea \label{sigma2} {
\vec{\sigma}}_\omega^R=\frac{1}{2}\kappa L_R \left[(B-B''){\bf
\hat{\Omega}}\wedge\left({\bf \hat{\Omega}}\wedge{\bf q}\right)+
B'{\bf \hat{\Omega}}\wedge {\bf q}-2B''\bf \hat{\Omega}\bf
\hat{\Omega}\cdot \bf q \right]. \label{2.4}\eea  \\
The interest to consider spatial distribution of vortices and
anisotropy of mutual friction in rotating container has led
Mathieu et al. in \cite{MPS2} to analyze a more general case in
which a parallelepipedic cavity filled of helium II rotates around
an axis tilted an angle $\theta$ with respect to its wall. In the
following Subsection we will show that the results of the latter
experiments can be easily explained using the
general expression (\ref{sigma2}). \\
Substituting the expression (\ref{2.4}) into the system (\ref{1})
and choosing ${\bf \Omega}=\left(\Omega,0,0\right)$, the system
assumes the following form:
\begin{equation}
\begin{cases}
\frac{\partial \rho}{\partial t}+\rho\frac{\partial v_j}{\partial
x_j }=0\cr \rho\frac{\partial v_i}{\partial t}+\frac{\partial
p}{\partial x_i}+2\rho \Omega v_j\epsilon_{1ji}=0\cr \frac{\partial
T}{\partial t}+\frac{1}{\rho c_V}\frac{\partial q_j}{\partial
x_j}=0\cr \frac{\partial q_i}{\partial t}+\zeta\frac{\partial
T}{\partial x_i}+\left(2\Omega-\frac{1}{2}B' \kappa L_R \right)
q_j\epsilon_{1ji}=\frac{1}{2}\kappa
L_R[(B-B'')\left(-q_i+q_1\delta_{i1}\right)-2B'' q_1 \delta_{i1}],
\end{cases}\label{1.1}
\end{equation}
where $\epsilon_{kji}$ is the Ricci tensor.
\par
\noindent It is easily observed that a stationary solution  of
this system is: \be \rho=\rho_0,\quad\!\!\! {\bf v}={\bf
0},\quad\!\!\! T=T_0,\quad\!\!\!{\bf q}={\bf 0}. \ee
\par
In order to study the propagation of plane harmonic waves of small
amplitude \cite{Whitham},  we put $\Gamma=(\rho, v_i, T, q_i)$, and
we look for solutions of the linearized system of field equations
(\ref{1}) of the form:
\begin{equation}
{\Gamma=\Gamma_0+\tilde \Gamma e^{i(Kn_jx_j-\omega t)}}, \label{1s1}
\end{equation}
 where {$\Gamma_0=(\rho_0, 0, T_0,  0)$} denotes the unperturbed state,
{$\tilde \Gamma=\left(\tilde \rho, \tilde v_i, \tilde T, \tilde
q_i\right)$} small amplitudes whose products can be neglected,
{$K=k_r+ik_s$} is the wavenumber, {$\omega =\omega_r+i\omega_s$}
the frequency and {${\bf n}=(n_i)$}  the unit vector orthogonal to
the wave front. Along this paper we will assume that the
propagating waves do not affect the vortex tangle, i.e. that they
do not contribute to the production nor the destruction of
vortices. In other terms, the waves are used to explore a given
vortex tangle, without modifying it. If the wave amplitude is high
enough, it could yield new contributions to the tangle.
\par
In the following we assume that $\Omega$ is small, so that the
term ${\bf i}_0$ in (\ref{1}b) can be neglected. For the sake of
simplicity, the subscript $0$, which denotes quantities referring
to the unperturbed state $\Gamma_0$, will be dropped out.

\par
\subsection{First case: ${\bf n}$ parallel to ${\bf \Omega}$}
In this subsection we analyze the case in which the unit vector
${\bf n}$ orthogonal to the wave front is parallel to the axis of
rotation, i.e. {${\bf n}=(1, 0, 0)$}. Substituting (\ref{1s1})
into the linearized system (\ref{1.1}) and letting {${\bf
t}_1=(0,0,1)$} and {${\bf t}_2=(0,1,0)$} as unit vectors tangent
to the wave front, the following homogeneous algebraic linear
system for the small amplitudes  is obtained:
\begin{equation}\label{rotnpo}
\begin{cases}
-\omega \tilde \rho + \rho K \tilde v_1=0 \cr -\omega\tilde v_1
+K\frac{p_\rho}{\rho} \tilde \rho=0
 \cr
 -\omega \tilde T+\frac{K}{\rho c_V}\tilde q_1=0
 \cr (-\omega-iB''\kappa L_R)\tilde q_1+\zeta K\tilde T =0
 \cr \cr -\omega\tilde
v_3-2i \Omega\tilde v_2=0 \cr -\omega\tilde v_2+2i\Omega\tilde v_3=0
\cr\cr

\left(-\omega-\frac{i}{2}\kappa L_R (B-B'')\right)\tilde
q_3-i\left(2\Omega-\frac{1}{2}\kappa L_R B'\right)\tilde q_2=0 \cr
\left(-\omega-\frac{i}{2}\kappa L_R(B-B'')\right)\tilde
q_2+i\left(2\Omega-\frac{1}{2}\kappa L_R B'\right)\tilde q_3=0.
\end{cases}
\end{equation}

From the above system, it follows that longitudinal and
transversal modes evolve independently. The study of the
longitudinal modes furnishes the existence of two waves: the first
is known as {\it first sound} or {\it pressure wave} in which
density and velocity vibrate, and the second is known as {\it
second sound} or {\it temperature wave} in which temperature and
heat flux vibrate. Therefore, as observed in \cite{Snyder}, when
the wave is propagated parallel to the rotation axis, the
longitudinal modes are influenced by the rotation only through the
axial component of the mutual friction ($B''$ coefficient). In
fact, the first two equations of the system (\ref{rotnpo}), give
for the first sound $V_1:=\frac{\omega}{K}= \sqrt{p_\rho}$,
whereas the third and fourth equation, with the assumption
$K=k_r+ik_s$ and $\omega$ real, give second sound waves with the
following velocity and attenuation: \be \label{3.8}
w_2:=\frac{\omega}{k_r}= \sqrt{\frac{4 V_2^4 k_r^2}{4 V_2^2
k_r^2+B''^2 \kappa L_R^2}}\quad \textrm{and} \quad k_s=\frac{w_2
B'' \kappa L_R}{2 V_2^2}\ee where $V_2^2:=\frac{\zeta}{\rho c_V}$
is the velocity of the second sound in the absence of vortices.
Therefore, the following fields vibrate respectively:
\begin{center}
\begin{tabular}{|l|l|}
{$\omega_{1,2}=\pm k V_1$}&{$\omega_{3,4}=\pm \sqrt{\frac{4 V_2^4
k_r^4}{4
V_2^2 k_r^2+B''^2 \kappa L_R^2}}$}  \\
\hline\hline &                          \\
{$\tilde \rho=\psi $}&{$\tilde \rho=0$}\\
{$\tilde v_1=\pm \frac{V_1}{\rho}\psi$}&{$\tilde v_1=0$}  \\
{$\tilde T_0=0  $ } &{$\tilde T=T_0\psi $}\\
{$\tilde q_1=0  $ } &{$\tilde q_1=\pm \rho c_V T_0 \sqrt{\frac{4
V_2^4 k_r^4}{4
V_2^2 k_r^2+B''^2 \kappa L_R^2}} \ \psi$}\\
 \end{tabular}
\end{center}

\par On the contrary, the transversal modes are influenced by the
rotation. In fact, by considering the fifth and the sixth equation
of (\ref{rotnpo}) they admit nontrivial solutions if and only if
its determinant vanishes; this yields $\omega_{5, 6}=\pm
2|\Omega|$.
\par
Now, we consider the equations seven and eight of the system
(\ref{rotnpo}) and, as above, we find the following dispersion
relation: \be \left(2\Omega-\frac{1}{2}\kappa L_R B'\right)^2-
\left(-\omega-\frac{i}{2}\kappa L_R (B-B'')\right)^2
=0,\label{rd0} \ee whose solutions are \be\label{3.16}
\omega_{7,8}=\pm (2\Omega-\frac{1}{2}\kappa L_R
B')-\frac{i}{2}\kappa L_R (B-B'').
 \ee
 These transversal modes are influenced from both
 dissipative and nondissipative contributions $B$, $B'$ and $B''$ in the
 interaction between quasi-particles and vortex lines.

\subsection{Second case: ${\bf n}$ orthogonal to ${\bf \Omega}$}
In this subsection we assume that the direction of propagation of
the waves is orthogonal to the rotation axis, i.e. for example,
${\bf n}=\left(0, 1, 0\right)$. The unit vectors tangent to the
wave front are ${\bf t}_1=\left(1,0,0\right)$ and ${\bf
t}_2=\left(0,0,1\right)$. Under these assumptions, substituting
(\ref{1s1}) into the linearized system (\ref{1.1}), the following
system is obtained:
\begin{equation}\label{rotnparo}
\begin{cases}
-\omega \tilde \rho + \rho K \tilde v_2=0 \cr -\omega\tilde v_2
+K\frac{p_\rho}{\rho} \tilde \rho+2i\Omega\tilde v_3=0
 \cr
 -\omega\tilde
v_3-2i\Omega\tilde v_2=0 \cr \cr
 -\omega \tilde T+\frac{K}{\rho c_V}\tilde q_2=0
 \cr \left(-\omega-\frac{i}{2}\kappa L_R (B-B'')\right)\tilde q_2+\zeta K\tilde
 T+
i(2\Omega-\frac{1 }{2}\kappa L_R B')\tilde q_3=0
\cr\left(-\omega-\frac{i}{2}\kappa L_R (B-B'')\right)\tilde
q_3-i\left(2\Omega-\frac{1}{2}\kappa L_R B'\right)\tilde q_2=0\cr\cr
-\omega\tilde v_1=0\cr (-\omega-iB''\kappa L_R)\tilde
q_1=0.\end{cases}
\end{equation}
In this case, the longitudinal and transversal modes do not evolve
independently. The first sound is coupled with one of the two
transversal modes in which velocity vibrates; while the second
sound is coupled with a transversal mode in which heat flux
vibrates.

Studying the first three equations of the system (\ref{rotnparo}),
we obtain a dispersion relation whose solutions are: \bea \omega_1&=&0,\label{sol2b}\\
w_{2,3}&=& \pm
V_1\sqrt{\left(1-4\frac{\Omega^2}{\omega_{2,3}^2}\right)^{-1}}.\label{sol2}
\eea  Summarizing:
\begin{center}
\begin{tabular}{|l|l|}
{$\omega_{1}=0$ }& {$\omega_{2,3}\simeq\pm K V_1+O(\Omega^2)$} \\
\hline\hline &
                            \\
{$\tilde \rho=\psi $} & {$\tilde \rho=\psi $}\\
{$\tilde v_2=0$} & {$\tilde v_2=\frac{\pm V_1}{\rho}\psi$}\\
{$\tilde v_3=i\frac{K V_1^2}{2\Omega\rho}\psi$} & {$\tilde v_3=-\frac{2i\Omega}{\rho K}\psi$}\\
 \end{tabular}
\end{center}
The second three equations admit non trivial solutions if and only
if their determinant vanishes. Neglecting the second-order terms
in $\Omega$, the dispersion relation becomes: \be \label{relaz}
\left(-\omega-\frac{i}{2}\kappa L_R
(B-B'')\right)\left[-\omega\left(-\omega-\frac{i}{2}\kappa
L_R(B-B'')\right)-K^2V_2^2\right]=0 \ee For $\omega\in \Re$ and
$K=k_r+ik_s$ complex, one gets the solution $\omega_4=0$, which
represents  a stationary mode; and two solutions which furnish the
following phase velocity and attenuation coefficient of the
temperature wave: \bea && w_2^2:=\frac{\omega^2}{k_r^2}=
V_2^2\frac{2}{1+\sqrt{1+\frac{(B-B'')^2\kappa^2 L_R^2}{4\omega^2}}},\\
&& k_s= \frac{(B-B'')\kappa L_R w_2}{4V_2^2}.\eea The approximated
solutions to second order in $\frac{(B-B'')\kappa L_R}{\omega}$
are: \bea &&
w_2\simeq V_2\left(1-\frac{(B-B'')^2\kappa^2 L_R^2}{32\omega^2}\right)+O\left(\frac{(B-B'')^4\kappa^4 L_R^4}{\omega^4}\right),\\
&& k_s\simeq \frac{(B-B'')\kappa
L_R}{4V_2}+O\left(\frac{(B-B'')^3\kappa^3 L_R^3}{\omega^2}\right)
\eea

Summarizing, when the direction of propagation of the waves is
orthogonal to the rotation axis, the temperature wave experiences
a strong attenuation, which grows with $\Omega$. The corresponding
modes are:
\begin{center}
\begin{tabular}{|l|l|}
$\omega_{4}=0$& {$\omega_{5,6}\simeq\pm k_r V_2\left(1-\frac{(B-B'')^2\kappa^2 L_R^2}{32\omega^2}\right)$}  \\
\hline\hline
                           & \\
{$\tilde T=-\frac{i(2\Omega-\frac{1}{2}\kappa L_R B')}{\zeta K}\psi$}& {$\tilde T=T_0\psi $}\\
{$\tilde q_2=0$} & {$\tilde q_2=\frac{T_0\zeta}{V_2}\left(1-\frac{(B-B'')^2\kappa^2 L_R^2}{32\omega^2}\right)\psi$}\\
{$\tilde q_3=\psi$} & {$\tilde
q_3=\frac{i\left(2\Omega-\frac{1}{2}\kappa L_R B'\right)
T_0\zeta\left(1-\frac{(B-B'')^2\kappa^2 L_R^2}{32\omega^2}\right)}
{V_2\left[\pm k_r V_2\left(1-\frac{(B-B'')^2\kappa^2 L_R^2}{32\omega^2}\right)-\frac{i}{2}(B-B'')\kappa L_R\right]}\psi$}\\
 \end{tabular}
\end{center}
We note that in the mode $\omega_4=0$, only
the transversal component of the heat flux is involved.\\
\noindent For $\omega=\omega_r+i\omega_s$ complex and $K\in \Re$,
the solutions of dispersion relation (\ref{relaz}) are:
\begin{eqnarray*} \omega_4
&=&-\frac{i}{2}(B-B'')\kappa L_R,\\
\omega_{5,6} &=& \pm
\sqrt{K^2V_2^2-\frac{1}{16}(B-B'')^2\kappa^2L_R^2}-i\frac{(B-B'')\kappa
L_R}{4}.
\end{eqnarray*}
The first mode, with $\omega_4=-\frac{i}{2}(B-B'')\kappa L_R$,
corresponds to an extremely slow relaxation phenomenon involving the
temperature wave and the transversal component of the heat flux:
\begin{center}
\begin{tabular}{|l|}
{$\omega_{4}=-\frac{i}{2}(B-B'')\kappa L_R$} \\
\hline\hline
                            \\
{$\tilde T=-\frac{i(2\Omega-\frac{1}{2}\kappa L_R B')}{\zeta K}\psi$}\\
{$\tilde q_2=0$}\\
{$\tilde q_3=\psi$}\\
 \end{tabular}
\end{center}
which when $\Omega\rightarrow 0$, converges to a stationary mode.
The attenuation in $\omega_{5,6}$ (corresponding to $q_1$ and
$q_2$) is physically reasonable in view of (\ref{2.4}), where it
is seen that for ${\bf q}$ parallel to ${\bf \Omega}$ the only
component of the friction force is the axial one (related to the
coefficient $B''$), whereas for ${\bf q}$ orthogonal to the vortex
line (i.e. to ${\bf \Omega}$) there is an attenuation dependent on
the dissipative coefficient ($B-B''$).

\section{Wave propagation in presence of thermal counterflow}
\setcounter{equation}{0}  In this section, we study wave
propagation in presence of pure thermal counterflow in liquid
Helium II to compare the results with those of \cite{JLM} and
those of Section 5. Let us consider a flow channel that connects
two He II reservoirs (as shown in fig. 1). When a steady heat is
applied to one end of the channel, there exists a temperature
difference $\Delta T$ between the two ends. From the microscopic
point of view using the two-fluid model, since only the normal
fluid component carries entropy and heat flow, it will move away
from the heat source (left reservoir) to the right reservoir and
then give up the heat. At the same time, the superfluid component
must counter-flow from right to left to conserve the mass. When it
arrives at the left reservoir, part of the superfluid component
will be converted to normal fluid by absorbing heat. Thus, a
relative counterflow between the normal fluid and superfluid
components is established, and this internal convection process is
termed thermal counterflow, which is associated to the heat flux
${\bf q}$ through the relation ${\bf q}=\rho_s T s \bf V$.

\par\noindent In this case, assuming that the vortex tangle caused
by the counterflow is isotropic, the vorticity tensor ${\bf
P}_\omega$, as indicated in \cite{JLM}, takes the following form:
 \bea\label{4.1} {\bf
P}_\omega^H= \frac{1}{3} \kappa B L {\bf U} \ \ \ \ \Rightarrow \
\ \ \ {\vec{\sigma}}_\omega^H=-\frac{1}{3}\kappa B L{\bf q}, \eea
where $L=\gamma^2q^2$. Under this assumption, the linearized set
of field equations read as:
\begin{equation}\label{sist}
\begin{cases}
\frac{\partial \rho}{\partial t}+\rho\frac{\partial v_j}{\partial
x_j }=0\cr \rho\frac{\partial v_i}{\partial t}+\frac{\partial
p}{\partial x_i}=0\cr \frac{\partial T}{\partial t}+\frac{1}{\rho
c_V}\frac{\partial q_j}{\partial x_j}=0\cr \frac{\partial
q_i}{\partial t}+\zeta\frac{\partial T}{\partial
x_i}=-\frac{1}{3}\kappa B Lq_i
\end{cases}
\end{equation}

A stationary solution of the system (\ref{sist}) is \cite{JLM}:
\be \rho=\rho_0,\quad\!\!\! \dot{\bf v}={\bf 0},\quad\!\!\!
T=T(x)=T_0-\frac{\kappa B L}{3\zeta}q_0 x,\quad\!\!\!{\bf q}={\bf
q}_0 \ee where $x$ is the direction of the heat flux ${\bf q}={\bf
q}_0$. In order to study the propagation of harmonic plane waves
in the channel, we look for solutions of the system (\ref{sist})
of the form:
\begin{equation}
{\Gamma=\Gamma_0+\tilde \Gamma e^{i(Kn_jx_j-\omega t)}}, \label{1s}
\end{equation}
 where {$\Gamma_0=(\rho_0, 0, T(x), {\bf q}_0)$}, and
the following homogeneous algebraic linear system for the small
amplitudes is obtained:
\begin{equation}
\begin{cases}
-\omega\tilde\rho+\rho K\tilde{v}_jn_j=0\cr
-\rho\omega\tilde{v}_i+p_\rho K\tilde\rho n_i=0\cr -\omega\tilde T
+\frac{K}{\rho c_V}\tilde q_jn_j=0\cr
\left(\omega+\frac{1}{3}i\kappa B L\right)\tilde q_i-\zeta K\tilde T
n_i=0.
\end{cases}\label{2b}
\end{equation}
The longitudinal modes are obtained projecting the vectorial
equations for the small amplitudes of velocity and heat flux on
the direction orthogonal to the wave front. It is observed that
the first sound is not influenced by the thermal counterflow,
while the velocity and the attenuation of the second sound are
influenced by the presence of the vortex tangle. The results are:
\[w_1 = \pm
\sqrt{p_\rho}
\]
with $p_\rho$ standing for $\pard p/\pard \rho $ and: \beas
w_2&=&V_2\sqrt{\left(1+\frac{k_s^2V_2^2}{\omega^2}\right)^{-1}}
\quad\Rightarrow\quad w_2\simeq
V_2\left(1-k_s^2\frac{V_2^2}{2\omega^2}\right),\\
k_s&=&\frac{1}{6}\kappa B L w_2.\eeas These results generalize
those of \cite{JLM} where the terms in $k_s^2$ have been
neglected.
\par
The transversal modes are obtained projecting the vectorial
equations for the small amplitudes of velocity and heat flux on the
wave front, obtaining:
\begin{equation}
\begin{cases}
-\omega\tilde{v}_\pi=0\cr \left(-\omega+\frac{i}{3}\kappa B
L\right)\tilde q_\pi=0
\end{cases}\label{2c}
\end{equation}
where $\pi$ denotes the tangential plane to the wave front. The
solutions of this equation are: $\omega_{1}=0$ and
$\omega_2=\frac{i}{3}\kappa B L$. The first mode ($\omega_{1}=0$)
is a stationary mode.

\section{Wave  propagation with simultaneous rotation and counterflow}
\setcounter{equation}{0} The combined situation  of rotation and
heat flux (as shown in fig. 2), is a relatively new area of
research \cite{JM04}-\cite{Tsubota2004}. The first motivation of
this great interest is that from the experimental observations one
deduces that the two effects are not merely additive; in
particular, for ${\bf q}$ or ${\bf \Omega}$ high, the measured
values of $L$ are always less than $L_H+L_R $.

Under the simultaneous influence of thermal counterflow {\bf $V$}
and rotation speed {\bf $\Omega$}, rotation produces an ordered
array of vortex lines parallel to rotation axis, whereas
counterflow velocity causes a disordered tangle. In this way the
total vortex line is given by the superposition of both
contributions so that the vortex tangle is anisotropic
\cite{JM05}, \cite{JMon}. Therefore, assuming that the rotation is
along the $x$ direction ${\bf \Omega}=\left(\Omega, 0, 0\right)$
and isotropy in the transversal $(y-z)$ plane, for the vorticity
tensor ${\bf P}_\omega$, in combined situation of counterflow and
rotation, the following explicit expression is taken:
 \be \label{pesp2} {\bf P}_\omega=\gamma \kappa
L\left\{\frac{2}{3}(1-D){\bf U}+D
\left[\left(1-\frac{B''}{B}\right)\left({\bf U}-{\bf
\hat{\Omega}}{\bf \hat{\Omega}}
 \right)+\frac{B'}{B}{\bf W}\cdot{\bf \hat{\Omega}}+2\frac{B''}{B}{\bf \hat{\Omega}}{\bf \hat{\Omega}}\right]\right\} \ee
where $\gamma$ is linked to the coefficient $B$ through the
relation $\gamma=B/2$ and $D$ is a parameter between $0$ and $1$
related to the anisotropy of vortex lines, describing the relative
weight of the array of vortex lines parallel to $\bf \Omega$ and
the disordered tangle of counterflow (when $D=0$ we recover an
isotropic tangle -- Eq. (\ref{4.1}), whereas when $D=1$ the
ordered array -- Eq. (\ref{Pomega})). Assuming
$b=\frac{1}{3}(1-D)+\frac{DB''}{B}$ and $c=\frac{B'D}{B}$, the
vorticity tensor (\ref{pesp2}) can be written as: \be\label{5.38}
{\bf P}_\omega=\gamma \kappa L\left\{\left(\begin{matrix}
2b&0&0\\
0&1-b&0\\
0&0&1-b
\end{matrix}\right)+\left(\begin{matrix}
0&0&0\\
0&0&c\\
0&-c&0
\end{matrix}\right)\right\}. \ee
Note that the isotropy in the $y-z$ plane may only be assumed when
both ${\bf \Omega}$ and ${\bf V}$ are directed along the $x$ axis. A
more general situations is yet an open topic.
\par\noindent Substituting the expression
(\ref{5.38}) into the linearized set of field equations (\ref{1}),
it assumes the following form:
\begin{equation}\label{siste1}
\begin{cases}
\frac{\partial \rho}{\partial t}+\rho\frac{\partial v_j}{\partial
x_j }=0\cr \rho\frac{\partial v_i}{\partial t}+\frac{\partial
p}{\partial x_i}+2\rho\Omega v_j\epsilon_{1ji}=0\cr \frac{\partial
T}{\partial t}+\frac{1}{\rho c_V}\frac{\partial q_j}{\partial
x_j}=0\cr \frac{\partial q_i}{\partial t}+\zeta\frac{\partial
T}{\partial x_i}+2\Omega q_j\epsilon_{1ji}=-\gamma\kappa L\left\{2b
q_1\delta_{1i}+\left[\left(1-b\right)q_2+cq_3\right]\delta_{2i}+
\left[\left(1-b\right)q_3-cq_2\right]\delta_{3i}\right\}
\end{cases}
\end{equation}
A stationary solution of this system is: \beas
&&\rho=\rho_0,\quad\!\!\! \dot{\bf v}={\bf 0},\quad\!\!\!{\bf
q}={\bf q}_0\equiv\left(q_{0}, 0, 0\right)\\
&&\!\!\!\!\!\!\!T=T(x_i)=T_0-2\frac{\gamma\kappa
L}{\zeta}bq_{0}\delta_{1i}x_i. \eeas
\par In order to study the propagation of harmonic plane waves,
we look for solutions of (\ref{siste1}) of the following form:
\begin{equation}
{\Gamma=\Gamma_0+\tilde \Gamma e^{i(Kn_jx_j-\omega t)}}, \label{1s}
\end{equation}
 where {$\Gamma_0=(\rho_0,\quad\!\!\!\!\! 0,\quad\!\!\!\!\! T(x_i),\quad\!\!\!\!\! {\bf
 q}_0)$} and $T(x_i)$ is a linear function of $x_i$.

Now, we investigate two different cases: ${\bf n}$ parallel to
${\bf \Omega}$ and ${\bf n}$ orthogonal to ${\bf \Omega}$; the
latter is the only case for which experimental data exist
\cite{SBD}.

\subsection{First case: ${\bf n}$ parallel to ${\bf
\Omega}$} In this subsection we analyze the case in which the unit
vector ${\bf n}$ orthogonal to the wave front is parallel to the
direction of the rotation, i.e. {${\bf n}=(1, 0, 0)$}. Letting
{${\bf t}_1=(0,1,0)$} and {${\bf t}_2=(0,0,1)$} as unit vectors
tangent to the wave front, the system (\ref{siste1}) for the small
amplitudes (\ref{1s}) is: \be
\begin{cases}
-\omega\tilde\rho+K\rho\tilde v_1=0\cr -\omega\tilde
v_1+K\frac{p_\rho}{\rho}\tilde\rho=0\cr -\omega\tilde
T+\frac{K}{\rho c_V}\tilde q_1=0\cr \left[-\omega -2i\gamma \kappa
Lb\right]\tilde q_1 +\zeta K\tilde T=0\cr \cr -\omega\tilde
v_2+2i\Omega\tilde v_3=0 \cr -\omega\tilde v_3-2i\Omega\tilde v_2=0
\cr\cr \left[-\omega-i\gamma \kappa L\left(1-b\right)\right]\tilde
q_2+\left(2i\Omega -i\gamma\kappa L c\right)\tilde q_3=0
 \cr
\left[-\omega-i\gamma \kappa L\left(1-b\right)\right]\tilde
q_3-\left(2i\Omega -i\gamma\kappa L c\right)\tilde q_2=0\label{sc1}
\end{cases}
\ee In this case the longitudinal and transversal modes evolve
independently. In particular, we can observe that the first sound,
given by the study of the first two equations of the system
(\ref{sc1}), is not influenced by the presence of the vortex tangle:
\begin{center}
\begin{tabular}{|l|}
{$\omega_{1,2}=\pm k_r V_1$} \\
\hline\hline
                            \\
{$\tilde \rho=\psi$}\\
{$\tilde v_1=\frac{V_1}{\rho}\psi$}
 \end{tabular}
\end{center}
whereas the second sound suffers extra attenuation due to the
vortex tangle. The third and fourth equation of the system
(\ref{sc1}) admit non trivial solutions if and only if their
determinant vanishes, obtaining in this way the following
dispersion relation: \be \omega^2+2i\gamma\kappa
Lb\omega-K^2V_2^2=0. \ee Supposing that $\omega$ is real and
$K=k_r+ik_s$ is complex, the dispersion relation admits the
solutions: \bea && w_2^2:=\frac{\omega^2}{k_r^2}=
V_2^2\frac{2}{1+\sqrt{1+\frac{4\gamma^2\kappa^2L^2b^2}{\omega^2}}},\label{5.45}\\
&& k_s=\frac{\gamma\kappa Lbw_2}{V_2^2}\label{5.46}.\eea When
$\Omega=0$ and $b=1/3$ the results of the Section 4 are obtained.
The approximate solutions to second order in $\frac{\gamma\kappa
Lb}{\omega}$ are: \bea &&
w_2\simeq V_2\left(1-\frac{\gamma^2\kappa^2L^2b^2}{2\omega^2}\right)+O\left(\frac{\gamma^4\kappa^4L^4b^4}{\omega^4}\right),\label{5.47}\\
&& k_s\simeq\frac{\gamma\kappa
Lb}{V_2}+O\left(\frac{\gamma^3\kappa^3L^3b^3}{\omega^2}\right)\label{5.48}.
\eea

Now, we study the transversal modes. The second subsystem (fifth
and sixth equation) of the system (\ref{sc1}) admits nontrivial
solutions if and only if its determinant vanishes; this yields:
\begin{equation}
\omega^2-4\Omega^2=0.
\end{equation}
The solutions of this equation are $\omega_{5, 6}=\pm 2|\Omega|$.
The respective modes are:
\begin{center}
\begin{tabular}{|l|}
{$\omega_{5,6}=\pm 2|\Omega|$} \\
\hline\hline
                            \\
{$\tilde v_3=\psi$}\\
{$\tilde v_2=\pm i\psi$}
 \end{tabular}
\end{center}
and they correspond to extremely slow phenomena, which, when
$\Omega\rightarrow 0$, tend to stationary modes.
\par
Finally, we consider the last subsystem (equations seven and
eight), whose dispersion relation is: \be \omega^2+2i\gamma\kappa
L(1-b)\omega +\left[-\left(\gamma\kappa L(1-b) \right)^2
-4\Omega^2+4\gamma\Omega\kappa L c-(\gamma\kappa L c)^2\right]=0
\ee which admits the following exact solutions: \be
\omega_{7,8}=\pm \left(2\Omega-\gamma\kappa
Lc\right)-i\gamma\kappa L\left(1-b\right).\label{5.52} \ee The
corresponding modes are:
\begin{center}
\begin{tabular}{|l|}
{$\omega_{7,8}=\pm \left(2\Omega-\gamma\kappa
Lc\right)-i\gamma\kappa L\left(1-b\right)$} \\
\hline\hline
                            \\
{$\tilde q_3=\psi$}\\
{$\tilde q_2=\pm i\psi$}
 \end{tabular}
\end{center}
From (\ref{5.45}), (\ref{5.46}) and (\ref{5.52}) one may obtain
the following quantities $L$, $b$ and $c$: \be \label{casoort}
L=\frac{-\omega_sw_2+V_2^2k_s}{\gamma\kappa w_2}, \quad\!\!\!
b=\frac{V_2^2k_s}{-\omega_sw_2+V_2^2k_s},\quad\!\!\!
c=\frac{-\omega_rw_2+2\Omega w_2}{-\omega_s w_2+V_2^2k_s}\ee where
we have put $\omega_7=\omega_r+i\omega_s$. \\ The results of this
section, from the physical point of view, imply that measurement
in a single direction are enough to give information on all the
variables describing the vortex tangle.

\subsection{Second case: ${\bf n}$ orthogonal to ${\bf \Omega}$}
 Now we assume that the direction of propagation of the waves is
orthogonal to the rotation axis, i.e. for example, ${\bf n}=\left(0,
1, 0\right)$. The unit vectors tangent to the wave front are ${\bf
t}_1=\left(1,0,0\right)$ and ${\bf t}_2=\left(0,0,1\right)$. Under
these assumptions, the homogeneous algebraic linear system for the
small amplitudes  is: \be
\begin{cases}
-\omega\tilde\rho+K\rho\tilde v_2=0\cr -\omega\tilde
v_2+K\frac{p_\rho}{\rho}\tilde \rho+2i\Omega\tilde v_3=0\cr -\omega
\tilde v_3-2i\Omega\tilde v_2=0\cr\cr -\omega\tilde T+\frac{K}{\rho
c_V}\tilde q_2=0\cr -\omega \tilde q_2+\zeta K\tilde
T+2i\Omega\tilde q_3=i\gamma \kappa L\left[(1-b)\tilde q_2+c\tilde
q_3\right] \cr
 -\omega\tilde
q_3-2i\Omega\tilde q_2=i\gamma \kappa L\left[(1-b)\tilde q_3-c\tilde
q_2\right]\cr \cr -\omega\tilde v_1=0\cr \left[-\omega-i\gamma
\kappa 2Lb\right]\tilde q_1=0 \label{src2}
\end{cases}
\ee In this case the longitudinal and the transversal modes not
evolve independently. In particular, the first sound is coupled with
one of the two transversal modes in which velocity vibrates, while
the second sound is coupled with a transversal mode in which heat
flux vibrates.

\par \noindent As in the previous subsection,
the first subsystem (first three equations) of the system
(\ref{src2}),   admits non trivial solutions if and only if its
determinant vanishes: \be
-\omega\left[\omega^2-4\Omega^2-K^2p_\rho \right]=0.\ee The
solutions of this equation, as also the corresponding modes, are
the same to the case of pure rotation (see equations
(\ref{sol2b})-(\ref{sol2})).

The second subsystem (fourth and fifth equations), has the
dispersion relation: \be \left(-\omega-i\gamma \kappa
L(1-b)\right)\left[\omega\left(-\omega-i\gamma \kappa
L(1-b)\right)+K^2 V_2^2\right]+\omega\left(2i\Omega-i\gamma \kappa
Lc\right)^2=0. \ee Assuming $\omega\in\Re$ and $K=k_r+ik_s$, one
obtains the following two equations: \bea
&&-\omega^3+\gamma^2\kappa^2
L^2(1-b)^2\omega+4\Omega^2\omega+\gamma^2\kappa^2L^2c^2\omega-4\gamma\kappa
L c \Omega\omega+\nonumber\\
&&\hspace{2.5cm} +k_r^2V_2^2\omega-k_s^2V_2^2\omega-2\gamma\kappa
L(1-b)k_rk_sV_2^2=0 ,\label{sol4b} \\
&& -2\gamma \kappa L(1-b)\omega^2+2k_rk_sV_2^2\omega+\gamma\kappa
L (1-b)(k_r^2-k_s^2)V_2^2=0.\label{sol3}\eea In the hypothesis of
small dissipation ($k_r^2\gg k_s^2$), from (\ref{sol3}) one
obtains: \be k_s=\gamma \kappa L
(1-b)\left(\frac{2w_2^2-V_2^2}{2w_2V_2^2}\right),\label{5.22} \ee
which substituting in (\ref{sol4b}), yields: \be
\omega^4-\left[\left(2\Omega-\gamma\kappa L c
\right)^2-\gamma^2\kappa^2L^2(1-b)^2\right]\omega^2-k_r^2V_2^2\omega^2-\gamma^2\kappa^2L^2(1-b)^2V_2^2k_r^2=0.\label{5.60}
\ee Putting $\tilde A=-\left[\left(2\Omega-\gamma\kappa L c
\right)^2-\gamma^2\kappa^2L^2(1-b)^2\right]$ and $\tilde
B=-\gamma^2\kappa^2L^2(1-b)^2$ and taking into account that
$w_2=\frac{\omega}{k_r}$, the Eq. (\ref{5.60}) becomes: \be
w_2^2\left[w_2^2\left(1+\frac{\tilde
A}{\omega^2}\right)-V_2^2\left(1-\frac{\tilde
B}{\omega^2}\right)\right]=0 \ee whose solutions are: \bea
&& w_2^2=0,\quad\mbox{and}\quad\nonumber\\
&& w_2^2=V_2^2\frac{\left(\omega^2-\tilde
B\right)}{\left(\omega^2+\tilde A\right)}=
 V_2^2\frac{1}{1-\frac{\left(2\Omega-\gamma\kappa
 Lc\right)^2}{\omega^2+\gamma^2\kappa^2L^2(1-b)^2}}.\label{5.26}\eea
We can remark that the coefficients $\tilde A$ and $\tilde B$ are
negative and that $w_2^2\geq V_2^2$ because $\omega^2+\tilde A
\leq \omega^2-\tilde B$ and, in particular, $w_2^2= V_2^2$ for
$\Omega=\frac{\gamma \kappa Lc}{2}$. Now, studying the transversal
modes, i.e. the third subsystem (equations seventh and eighth), we
obtain $ \omega_7=0$,  which corresponds to a stationary mode,
and: \be \omega_8=-i\gamma \kappa 2Lb.\label{5.25} \ee
Summarizing, also in this case measurement in a single direction
are enough to given information on all the variables describing
the vortex tangle, namely $L$, $b$ and $c$, from equations
(\ref{5.22}), (\ref{5.26}) and (\ref{5.25}): \bea \label{casopara}
&&L=\frac{4k_sw_2V_2^2-\omega_s\left(2w_2-V_2^2\right)}{2\left(2w_2^2-V_2^2\right)\gamma\kappa},\nonumber \\
&&b=-\frac{\omega_s\left(2w_2^2-V_2^2\right)}{4k_sw_2V_2^2-\omega_s\left(2w_2-V_2^2\right)},\\
&&c=\frac{4\Omega(2w_2^2-V_2^2)-\sqrt{(1-V_2^2)(4k_r^2(2w_2^2-V_2^2)^2+16k_s^2V_2^4)}}
{4k_sw_2V_2^2-\omega_s(2w_2^2-V_2^2)}\nonumber \eea  where we have
put $\omega_8=i\omega_s$ and $\omega_s=2\gamma \kappa L  B$.

In this subsection we have analyzed wave propagation in the combined
situation of rotation and counterflow with the direction {\bf $n$}
orthogonal to {\bf $\Omega$}. In \cite{SBD} Swanson et al.
experimented the same situation, but they didn't represent the
attenuation neither the speed of the second sound but only the
  vortex line density $L$ as function of  {\bf
$\Omega$} and {\bf $V$}. Therefore, it is unknown how they plotted
these graphics, which the hypothesis were made and what was the
anisotropy considered. Instead, the results of these two
subsections allow to know the spatial distribution of the vortex
tangle simply by performing experiments on waves propagating
orthogonally to {\bf $\Omega$} (equations (\ref{casoort})) or
parallelly to {\bf $\Omega$} (equations (\ref{casopara})).

\section{Conclusions}
In this work we have studied the propagation of waves
(longitudinal density and temperature waves, longitudinal and
transversal velocity and heat waves) in turbulent superfluid
helium in the three situations: rotating frame, thermal
counterflow, and simultaneous
thermal counterflow and rotation.\\
From the physical point of view it is interesting to note that our
detailed analysis in Section 5 shows that, in contrast to which
one could intuitively expect, measurements in a single direction
are enough to give information on all the variables describing the
vortex tangle, namely $L$, $b$ and $c$, for instance, from one of
(\ref{5.45})-(\ref{5.46}) and (\ref{5.52}) or of
(\ref{5.22})-(\ref{5.26}) and (\ref{5.25}). This is not an
immediate intuitive result. Future analyses of work along this
direction could be, for instance, to consider that ${\bf \Omega}$
and ${\bf V}$ have arbitrary directions, i.e. that they are not
parallel to each other, in which case (\ref{5.38}) would not be
sufficient to describe the vortex tangle, because no isotropy in
the $y-z$ plane could be assumed.
\par
Another topic could be to assume that the external waves produce
vibrations in the vortex lines, without creating nor destroying
them. An example of that is the work of Barenghi et al.
\cite{BTMA2}. A more general possibility would be to consider that
nonlinear effects of the external waves create and destroy new
vortices. Yet another topic would be to consider what happens with
waves whose wavevector $\lambda$ become short enough to be
comparable with the average vortex separation, of the order
$L^{-1/2}$. In this case, one could study nonlocal effects in the
vortex \cite{MonJ05, MonJou05}. The first mentioned application
could be carried out within the existing physical model, at the
expenses of more cumbersome calculations. In contrast, the other
three applications need more progress in the basic physical
understanding of the problem.

\section*{Acknowledgments}
We acknowledge Prof. D. Jou (Departament de F\'{\i}sica, Universitat
Aut\`{o}noma de Barcelona) and Prof. M.S. Mongiov\`{\i}
(Dipartimento di Metodi e Modelli Matematici, Universit\`{a} di
Palermo) for the enlightening discussions useful to the deepening of
these arguments. This work is supported by MIUR of Italy under
"Azioni Integrate Italia-Spagna, anno 2005". R.A. P. is supported by
an "Assegno di ricerca dell'Istituto Nazionale di Alta Matematica
'{\it F. Severi}' (INdAM)" of Italy and M.S. is supported by an
"Assegno di ricerca MIUR" of Italy.

\newpage
\begin{center} \mbox{\includegraphics[height=5cm,width=8cm]
{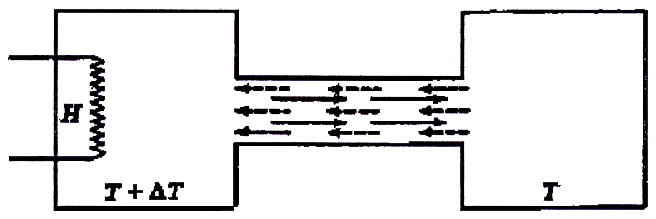}}\\
Fig. 1. Counterflow container configuration.
\end{center}

\begin{center} \mbox{\includegraphics[height=4cm,width=8cm]
{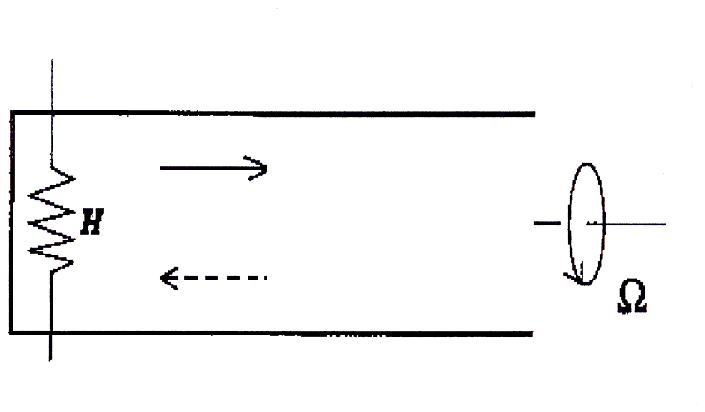}}
\\Fig. 2. Rotating counterflow container
configuration.
\end{center}


\begin{thebibliography}{99}

\bibitem{Tisza} L. Tisza, {Nature} {141} (1938) 913.

\bibitem{Landau} L.D. Landau, {J.Phys.} 5 (1941) 71.

\bibitem{Tough} J.T. Tough, Prog. Low Temp. Phys. 8 (1982)  133.

\bibitem{Donnelly}  R.J. Donnelly, Quantized Vortices in Helium II, Cambridge University Press, Cambridge, UK, 1991.

\bibitem{WSD} R.T. Wang, C.E. Swanson, R.J. Donnelly, Phys.Rev.B 36 (1987) 5240.

\bibitem{SBD} C.E Swanson, C.F. Barenghi, R.J. Donnelly, Phys.Rev.Lett. 50 (1983) 190.

\bibitem{M93} M.S. Mongiov\`{\i}, Phys. Rev. B 48 (1993) 6276.

\bibitem{JCL} D. Jou, J. Casas-V\'azquez, G. Lebon, Extended Irreversible Thermodynamics, Springer-Verlag, Berlin, 2001.

\bibitem{MR} I. M\"uller, T. Ruggeri, Rational Extended Thermodynamics, Springer-Verlag Berlin, 1998.

\bibitem{MPz} M.S. Mongiov\`{\i}, R.A. Peruzza, Zangew.Math.Phys. 54 (2003) 566.

\bibitem{MP04} M.S. Mongiov\`{\i}, R.A. Peruzza, I.J.Nonlinear Mech. 39 (2004) 1005.

\bibitem{MP03} M.S. Mongiov\`{\i}, R.A. Peruzza, Math. Comp. Model. 38 (2003) 409.

\bibitem{MPJ} M.S. Mongiov\`{\i}, R.A. Peruzza, D. Jou, Recent research developments in Physics, Transworld Research
Network, (2004), 1033.

\bibitem{JLM} D. Jou, G. Lebon, M.S. Mongiov\`{\i}, Phys.~Rev.~B 66 (2002) 224509.

\bibitem{BDV} C. F. Barenghi, R.J. Donnelly, W.F. Vinen, Quantized Vortex Dynamics and Superfluid Turbulence, Springer-Berlin, 2001.

\bibitem{JM04} D. Jou, M.S. Mongiov\`{\i}, Phys.~Rev.~B 69 (2004) 094513.

\bibitem{JM05} D. Jou, M.S. Mongiov\`{\i},  Phys.~Rev.~B 72 (2005) 144517.

\bibitem{JMon} D. Jou, M.S. Mongiov\`{\i}, Phys.~Rev.~B 74 (2006) 054509.

\bibitem{MJ05} M.S. Mongiov\`{\i}, D. Jou, Phys.~Rev.~B 72 (2005) 104515.

\bibitem{Tsubota04} M. Tsubota, C.F. Barenghi, T. Araki, A. Mitani, Phys.~Rev.~B 69 (2004) 134515.

\bibitem{Tsubota2004} M. Tsubota, T. Araki, C.F. Barenghi, Jour. Low Temp. Phys. 134 (2004) 471.

\bibitem{Mongiovi} M.S.~Mongiov\`{\i}, Physica~A, 291, (2001) 518.

\bibitem{MonJou07} M.S.~Mongiov\`{\i}  D.~Jou, Phys.~Rev.~B, 75, (2007)
024507.

\bibitem{JMS} D.~Jou, M.S.~Mongiov\`{\i}, M.~Sciacca, "Vortex density waves
in a hydrodynamical model of superfluid turbulence",
Phys.~Lett.~A, in press

\bibitem{Henderson} K.L. Henderson and C.F. Barenghi, Theoret. Comput. Fluid Dynam. 18 (2004)
183.

\bibitem{HV56} H.E. Hall, W. F. Vinen, Proc.~R.~Soc.~London~A238 (1956) 215; Proc.~Roy.~Soc. ~A238 (1956) 204.

\bibitem{Snyder} H.A. Snyder, Physics of Fluids~6 (1963) 755.

\bibitem{BTMA2} C.F. Barenghi, M. Tsubota, A. Mitani, T. Araki, J. Low Temp. Phys. 134 (2004) 489.

\bibitem{MPS2} P. Mathieu, B. Pla\c{c}ais, Y. Simon, Phys. Rev. B~29 (1984) 2489.

\bibitem{Whitham} J.Whitham, {Linear and Nonlinear Waves}, New York, Wiley, 1974.


\bibitem{kala} I.M. Khalatnikov, An Introduction to the Theory of Superfluidity, New York, Benjamin, 1965.

\bibitem{SB2} Y.A. Sergeev, C.F. Barenghi, J. Low. Tem. Phys. 127 (2002) 203.

\bibitem{MonJ05} M.S. Mongiov\`{\i}, D. Jou, Phys.~Rev.~B 71 (2005) 094507.

\bibitem{MonJou05} M.S. Mongiov\`{\i}, D. Jou, J.~Phys:~Condens.~Matter 17 (2005) 4423.

\end{thebibliography}
\end{document}